\documentclass[prd,twocolumn,tightenlines,showpacs,nofootinbib,superscriptaddress]{revtex4-1} 

\usepackage{amsfonts}
\usepackage{dcolumn}
\usepackage{bm}
\usepackage{amsmath}
\usepackage{nicefrac}
\usepackage{amsthm}
\usepackage{amssymb}
\usepackage{graphicx}
\usepackage{color}
\usepackage{url}
\usepackage{cancel}
\usepackage{rotating}

\usepackage[colorlinks=true,linkcolor=black,citecolor=blue,urlcolor=blue,bookmarksopen]{hyperref}

\newcommand{\appropto}{\mathrel{\vcenter{
  \offinterlineskip\halign{\hfil$##$\cr
    \propto\cr\noalign{\kern2pt}\sim\cr\noalign{\kern-2pt}}}}}
\renewcommand{\v}[1]{\boldsymbol{#1}}		

\begin{document}

\title{Probing low-mass vector bosons with parity nonconservation and nuclear anapole moment measurements in atoms and molecules}

\date{\today}
\author{V.~A.~Dzuba} 
\affiliation{School of Physics, University of New South Wales, Sydney 2052, Australia}
\author{V.~V.~Flambaum} 
\affiliation{School of Physics, University of New South Wales, Sydney 2052, Australia}
\affiliation{Johannes Gutenberg University of Mainz, 55128 Mainz, Germany}
\author{Y.~V.~Stadnik} 
\affiliation{Johannes Gutenberg University of Mainz, 55128 Mainz, Germany}

\begin{abstract}
In the presence of P-violating interactions, the exchange of vector bosons between electrons and nucleons induces parity-nonconserving (PNC) effects in atoms and molecules, while the exchange of vector bosons between nucleons induces anapole moments of nuclei. 
We perform calculations of such vector-mediated PNC effects in Cs, Ba$^+$, Yb, Tl, Fr and Ra$^+$ using the same relativistic many-body approaches as in earlier calculations of standard-model PNC effects, but with the long-range operator of the weak interaction. 
We calculate nuclear anapole moments due to vector boson exchange using a simple nuclear model. 
From measured and predicted (within the standard model) values for the PNC amplitudes in Cs, Yb and Tl, as well as the nuclear anapole moment of $^{133}$Cs, we constrain the P-violating vector-pseudovector nucleon-electron and nucleon-proton interactions mediated by a generic vector boson of arbitrary mass. 
Our limits improve on existing bounds from other experiments by many orders of magnitude over a very large range of vector-boson masses. 
\end{abstract}


\maketitle

\textbf{Introduction.} --- 
The discovery that the parity symmetry (that is, the symmetry associated with the inversion of the spatial coordinates) is not conserved by the weak interaction \cite{Yang1956,Wu1957} was pivotal for the subsequent development of the standard model (SM) of particle physics, which to date remains the most successful description of elementary particles and their interactions. 
At the same time, the SM does not explain a number of important observed phenomena, such as dark matter, suggesting the existence of at least one new feebly-interacting particle beyond the SM. 

Atomic parity nonconservation (PNC) experiments provide a very powerful and relatively inexpensive test of the SM at low energies \cite{Khriplovich1991Book,Ginges2004Review,Roberts2015ReviewPNC}. 
Measurements and calculations (within the SM) of the Cs $6s$-$7s$ PNC amplitude have to date provided the most precise atomic test of the electroweak theory \cite{Bouchiat1974,Bouchiat1982,Sushkov1989,Wood1997,Johnson1992,Kozlov2001,Safronova2002,Ginges2002,Shabaev2005,Derevianko2009,Roberts2012} and also invaluable information on parity-violating interactions within the nucleus \cite{Zeldovich1957,Flambaum1980,Flambaum1984,Wood1997,Murray1997}. 
Investigations of atomic PNC phenomena have been applied to search for new vector bosons with masses greater than $100~\textrm{keV}$ \cite{Bouchiat1983,Marciano1990,Bouchiat2004,Marciano2012a,Marciano2012b}, as well as interactions of electrons and nucleons with bosonic dark matter and dark-energy-type fields \cite{Stadnik2014a,Stadnik2014b,Stadnik2014c}. 

In the present work, we investigate the manifestation of the exchange of a generic vector boson $Z'$ of arbitrary mass between atomic electrons and nucleons, in the presence of the following couplings \cite{Footnote1}: 
\begin{equation}
\label{general_formula_SP}
\mathcal{L}_\textrm{int} = Z'_\mu \sum_{f=e,p,n} \bar{f} \gamma^\mu \left(g_f^V + \gamma_5 g_f^A \right) f \, . 
\end{equation}
The \emph{P}-violating potential due to the exchange of a vector boson of mass $m_{Z'}$ between two fermions reads: 
\begin{equation}
\label{relativistic_potential_FULL}
V_{12}(r) =  \frac{g_1^A g_2^V}{4 \pi} \frac{e^{-m_{Z'} r}}{r} \gamma_5 \, , 
\end{equation}
where $r$ is the distance between the two fermions, the $\gamma$-matrix corresponds to fermion 1, and we have treated fermion 2 nonrelativistically. 
We introduce the shorthand notation $g_N^V \equiv (N g_n^V + Z g_p^V) / A$, where $N$ is the neutron number, $Z$ is the proton number, and $A = Z + N$ is the nucleon number. 

The \emph{P}-violating potential in Eq.~(\ref{relativistic_potential_FULL}) induces PNC effects in atoms and molecules, as well as nuclear anapole moments, by mixing states of opposite parity. 
We first calculate such vector-mediated atomic PNC effects using the same relativistic many-body approaches as in earlier calculations of standard-model PNC effects \cite{s-d_PNC2001,Ginges2002,s-d_PNC2011,Dzuba2011YbPNC}, but with the long-range operator (\ref{relativistic_potential_FULL}). 
We then calculate the nuclear anapole moments due to vector boson exchange using a simple nuclear model. 
We find that atomic PNC experiments improve on existing bounds on the interaction parameters in Eq.~(\ref{relativistic_potential_FULL}) by many orders of magnitude over a very large range of vector-boson masses, compared with previous experiments that looked for macroscopic forces associated with potential (\ref{relativistic_potential_FULL}), since phenomena that arise on atomic and sub-atomic length scales allow one to probe larger boson masses compared with experiments that probe phenomena on macroscopic length scales. 


\begin{table*}[t]
\centering
\caption{
Summary of calculations of atomic PNC amplitudes induced by interaction (\ref{relativistic_potential_FULL}) for various vector boson masses. 
The presented values for the atomic PNC amplitudes are in terms of the parameter $- Q_W/N = (2 \sqrt{2} A g_e^A g_N^V) / (N G_F  m_{Z'}^2) $ and in the units $i \times \textrm{atomic unit (a.u.)}$. 
}
\label{tab:table1}
\resizebox{\linewidth}{!}{
\begin{tabular}{ |c|c|c|c|c|c|c|c| }%
\hline
$m_{Z'}$~(eV) & Cs ($6s - 7s$)  & Ba$^+$ ($6s - 5d_{3/2}$) & Yb ($6s^2 ~ ^1S_0 - 6s5d ~ ^3D_1$)  &  Tl ($6s^2 6p_{1/2} - 6s^2 6p_{3/2}$)  & Fr ($7s - 8s$)  & Fr ($7s - 6d_{3/2}$)  &  Ra$^+$ ($7s - 6d_{3/2}$)  \\ \hline
$\infty$ & $8.9 \times 10^{-12}$ & $2.3 \times 10^{-11}$ & $1.1 \times 10^{-9}$ & $2.8 \times 10^{-10}$ & $1.5 \times 10^{-10}$ & $-5.3 \times 10^{-10}$ & $4.5 \times 10^{-10}$   \\ \hline
$10^9$ & $8.6 \times 10^{-12}$ & $2.2 \times 10^{-11}$ & $1.1 \times 10^{-9}$ & $2.7 \times 10^{-10}$ & $1.5 \times 10^{-10}$ & $-5.1 \times 10^{-10}$ & $4.4 \times 10^{-10}$   \\ \hline
$10^8$ & $8.3 \times 10^{-12}$ & $2.1 \times 10^{-11}$ & $1.0 \times 10^{-9}$ & $2.6 \times 10^{-10}$ & $1.5 \times 10^{-10}$ & $-4.9 \times 10^{-10}$ & $4.2 \times 10^{-10}$   \\ \hline
$10^7$ & $6.3 \times 10^{-12}$ & $1.6 \times 10^{-11}$ & $6.7 \times 10^{-10}$ & $1.5 \times 10^{-10}$ & $8.1 \times 10^{-11}$ & $-2.7 \times 10^{-10}$ & $2.3 \times 10^{-10}$   \\ \hline
$10^6$ & $2.9 \times 10^{-12}$ & $7.0 \times 10^{-12}$ & $2.3 \times 10^{-10}$ & $3.8 \times 10^{-11}$ & $1.7 \times 10^{-11}$ & $-5.8 \times 10^{-11}$ & $4.6 \times 10^{-11}$   \\ \hline
$10^5$ & $2.2 \times 10^{-13}$ & $2.0 \times 10^{-13}$ & $2.2 \times 10^{-11}$ & $1.4 \times 10^{-12}$ & $6.2 \times 10^{-13}$ & $-2.7 \times 10^{-12}$ & $3.9 \times 10^{-13}$   \\ \hline
$10^4$ & $3.0 \times 10^{-15}$ & $-3.3 \times 10^{-15}$ & $3.8 \times 10^{-13}$ & $1.6 \times 10^{-14}$ & $7.1 \times 10^{-15}$ & $-3.6 \times 10^{-14}$ & $-4.8 \times 10^{-15}$   \\ \hline
$10^3$ & $1.8 \times 10^{-17}$ & $4.2 \times 10^{-17}$ & $1.4 \times 10^{-15}$ & $1.3 \times 10^{-16}$ & $4.7 \times 10^{-17}$ & $-1.7 \times 10^{-16}$ & $1.1 \times 10^{-16}$   \\ \hline
$10^2$ & $1.1 \times 10^{-19}$ & $5.7 \times 10^{-19}$ & $9.1 \times 10^{-18}$ & $1.2 \times 10^{-18}$ & $3.5 \times 10^{-19}$ & $-1.0 \times 10^{-18}$ & $1.4 \times 10^{-18}$   \\ \hline
$10$     & $1.1 \times 10^{-21}$ & $5.7 \times 10^{-21}$ & $8.9 \times 10^{-20}$ & $1.2 \times 10^{-20}$ & $3.4 \times 10^{-21}$ & $-1.0 \times 10^{-20}$ & $1.4 \times 10^{-20}$   \\ \hline
\end{tabular}
}
\end{table*}

\textbf{PNC effects in atoms.} --- 
The PNC amplitude associated with interaction (\ref{relativistic_potential_FULL}) for the atomic transition $a \to b$ can be written as:
\begin{align}
\label{PNC_amplitude_defn}
E_\textrm{PNC}^{a \to b} = \sum_n &\left[ \frac{\left< a \left| V_{eN} \right| n \right> \left< n \left| H_{E1} \right| b \right> }{E_a - E_n} \right. \\ \notag
&+ \left. \frac{\left< a \left| H_{E1} \right| n \right> \left< n \left| V_{eN} \right| b \right> }{E_b - E_n} \right] \, , 
\end{align}
where $H_{E1}$ is the electric-dipole operator. 

We perform calculations of vector-mediated atomic PNC effects starting from the relativistic Hartree-Fock-Dirac method including electron core polarisation corrections calculated in the framework of the random-phase approximation (RPA) method. 
For atoms with one external electron, we also use the correlation potential method \cite{Dzuba1987corr,Dzuba1989corr,Sushkov1989,s-d_PNC2001,Ginges2002,s-d_PNC2011} to take into account the dominating correlation corrections. 
For Yb and Tl, we employ the combination of the configuration interaction (CI) and many-body perturbation theory (MBPT) methods, CI+MBPT \cite{Dzuba2011YbPNC}, in the $V^{N-2}$ approximation for Yb and $V^{N-3}$ approximation for Tl, treating them as two- and three-valence-electron systems, respectively. 

We summarise the results of our calculations in Table~\ref{tab:table1} and present limits on the vector-mediated electron-nucleon interaction, as defined in Eq.~(\ref{general_formula_SP}), in Table~\ref{tab:table2}. 
We note that when a high-mass vector boson is exchanged, the induced PNC amplitude has a very strong $Z$-dependence (the relevant matrix elements scale as $\propto A Z^2 K_\textrm{rel}$, where $K_\textrm{rel}$ is a relativistic factor \cite{Bouchiat1974}), whereas when a low-mass vector boson is exchanged, the induced PNC amplitude has a milder $Z$-dependence (in the semi-classical framework, the relevant matrix elements scale only as $\propto A$).

\textbf{Nuclear anapole moments.} --- 
The nuclear anapole moment $\v{a}$ is expressed through the electromagnetic current density $\v{j}(\v{r})$ as follows: 
\begin{equation}
\label{anapole_moment_defn}
\v{a} = - \pi \int d^3 r ~ r^2 \v{j}(\v{r}) \, .
\end{equation}

We are specifically interested in heavy nuclei ($A \gg 1$) with a single unpaired nucleon. 
For our calculations, we adopt the simple shell model of the nucleus, with a constant core density $\rho_\textrm{core} (r) = \rho_0$, and treat all nucleons nonrelativistically. 
In the nonrelativistic limit, the potential (\ref{relativistic_potential_FULL}) for the interaction of an external nucleon $N'$ with the core nucleons $N$ reads: 
\begin{equation}
\label{nonrelativistic_potential_FULL}
V_{N'N}(r) = - \sum_N \frac{g_{N'}^A g_N^V}{8 \pi m} \left\{ \v{\sigma} \cdot \v{p} ~, ~\frac{e^{-m_{Z'} r}}{r} \right\}  \, , 
\end{equation}
where $m$, $\v{\sigma}$ and $\v{p}$ are the mass, spin and momentum operator for the external nucleon, respectively. 
We consider the two limiting cases:~(i) $m_{Z'} \gg 1/r_0$, and (ii) $m_{Z'} \ll 1/R$, where $r_0 \approx 1.2~\textrm{fm}$ is a distance parameter related to the internucleon separation, and $R = A^{1/3} r_0$ is the radius of the nucleus. 

We determine the wavefunction of the external nucleon, in the presence of interaction (\ref{nonrelativistic_potential_FULL}), by applying perturbation theory, making use of the relation $\v{p} = im [H,\v{r}]$, where $H$ is the nonrelativistic nuclear Hamiltonian, and summing over all intermediate states with the aid of the completeness relation. 
For simplicity, we neglect the spin-orbit interaction, and in the limiting case $m_{Z'} \ll 1/R$, we also first average over the Yukawa part of the potential in Eq.~(\ref{nonrelativistic_potential_FULL}) before applying the relation $\v{p} = im [H,\v{r}]$. 
The resulting wavefunction of the external nucleon reads: 
\begin{equation}
\label{nucleon_wavefunction_perturbed1}
\psi(\v{r}) \approx \left[ 1 + i \frac{g^A_{N'} g^V_N }{m_{Z'}^2} \rho_0 ~ \v{\sigma} \cdot \v{r} \right] \psi_0(\v{r}) ~~\textrm{for $m_{Z'} \gg 1/r_0$} \, ,
\end{equation}
\begin{equation}
\label{nucleon_wavefunction_perturbed2}
\psi(\v{r}) \approx \left[ 1 + i g^A_{N'} g^V_N \frac{3 A }{10 \pi R} ~ \v{\sigma} \cdot \v{r} \right] \psi_0(\v{r}) ~~\textrm{for $m_{Z'} \ll 1/R$} \, ,
\end{equation}
where $\psi_0(\v{r})$ is the unperturbed wavefunction of the external nucleon.

\begin{table*}[t]
\centering
\caption{ 
Summary of derived limits on the combinations of parameters $g_e^A g_N^V/m_{Z'}^2$ for $m_{Z'} \gg Z \alpha m_e$, $g_e^A g_N^V$ for $m_{Z'} \ll 1/R_\textrm{atom}$, $g_p^A g_N^V/m_{Z'}^2$ for $m_{Z'} \gg 1/r_0$ ($r_0 \approx 1.2~\textrm{fm}$), and $g_p^A g_N^V$ for $m_{Z'} \ll 1/R_\textrm{nucl}$, from the consideration of vector-mediated \emph{P}-violating interactions in atoms. 
We have also summarised the experimentally measured and theoretically predicted (within the standard model) nuclear-spin-independent PNC amplitudes used in deriving the limits on the electron-nucleon interaction. 
}
\label{tab:table2}
\resizebox{\linewidth}{!}{
\begin{tabular}{ c|c|c|c|c|c|c }%
Atom & $E_\textrm{PNC}^{\textrm{exp}}~(i10^{-11} ~\textrm{a.u.})$ & $E_\textrm{PNC}^{\textrm{theor}}~(i10^{-11} ~\textrm{a.u.})$  & $|g_e^A g_N^V|/m_{Z'}^2$ limit $(\textrm{GeV}^{-2})$ & $|g_e^A g_N^V|$ limit & $|g_p^A g_N^V|/m_{Z'}^2$ limit $(\textrm{GeV}^{-2})$ & $|g_p^A g_N^V|$ limit  \\ \hline 
$^{133}$Cs & $0.8353(29)$ \cite{Wood1997} & $0.8428(38)$ \cite{Roberts2012} & $3.9 \times 10^{-8}$ & $3.1 \times 10^{-14}$ & $2.3 \times 10^{-5}$ & $6.0 \times 10^{-8}$ \\ \hline
$^{174}$Yb & $87(14)$ \cite{Tsigutkin2009YbPNC} & $110(14)$ \cite{Dzuba2011YbPNC} & $1.1 \times 10^{-6}$ & $1.4 \times 10^{-12}$ & --- & --- \\ \hline 
$^{205}$Tl & 24.8(2) \cite{Fortson1995TlPNC} & 25.6(7) \cite{Kozlov2001TlPNC} & $1.5 \times 10^{-7}$ & $3.6 \times 10^{-13}$ & --- & --- \\ 
\end{tabular}
}
\end{table*}

We compute the anapole moment of a nucleus using Eqs.~(\ref{nonrelativistic_potential_FULL}), (\ref{nucleon_wavefunction_perturbed1}) and (\ref{nucleon_wavefunction_perturbed2}): 
\begin{align}
\label{anapole_moment1}
\v{a} &\approx - \frac{g^A_{N'} g^V_N}{m_{Z'}^2} \frac{2 \pi e \mu }{m } \rho_0 \left< r^2 \right> \frac{K \v{I}}{I(I+1)} \\ \notag 
&\approx - \frac{g^A_{N'} g^V_N}{m_{Z'}^2} \frac{9 e \mu }{10 m r_0 } A^{2/3} \frac{K \v{I}}{I(I+1)} ~~\textrm{for $m_{Z'} \gg 1/r_0$} \, ,
\end{align}
\begin{align}
\label{anapole_moment2}
\v{a} &\approx - g^A_{N'} g^V_N \frac{3 e \mu  }{5 m } \frac{A \left< r^2 \right>}{R} \frac{K \v{I}}{I(I+1)} \\ \notag 
&\approx - g^A_{N'} g^V_N \frac{9 e \mu r_0 }{25 m} A^{4/3} \frac{K \v{I}}{I(I+1)} ~~\textrm{for $m_{Z'} \ll 1/R$} \, ,
\end{align}
where $-e$ is the electric charge of the electron, $\mu$ is the magnetic moment of the external nucleon in nuclear magnetons ($\mu_p = 2.79$, $\mu_n = -1.91$), $\v{I}$ is the spin of the nucleus, and $K = (I+1/2) (-1)^{I+1/2-l}$, with $l$ being the orbital angular momentum of the external nucleon. 
In the second lines of Eqs.~(\ref{anapole_moment1}) and (\ref{anapole_moment2}), we have made use of the relations $\left< r^2 \right> \approx 3 r_0^2 A^{2/3} / 5$ and $\rho_0 = (4 \pi r_0^3 / 3)^{-1}$. 

The Hamiltonian for the interaction of atomic electrons with the electromagnetic vector potential of the nucleus created by an anapole moment takes the form:
\begin{equation}
\label{anapole_EM_Hamiltonian}
H_\textrm{anapole} = e \v{\alpha} \cdot \v{a} \delta(\v{r}) = \frac{G_F}{\sqrt{2}} \frac{K \v{I} \cdot \v{\alpha}}{I(I+1)} \kappa_a \delta(\v{r}) \, ,
\end{equation}
where $\v{\alpha} = \bigl( \begin{smallmatrix}0 & \v{\sigma}\\ \v{\sigma} & 0\end{smallmatrix}\bigr)$ are Dirac matrices associated with the external nucleon, and $G_F \approx 1.166 \times 10^{-5}~\textrm{GeV}^{-2}$ is the Fermi constant of the weak interaction. 
From Eqs.~(\ref{anapole_moment1}) and (\ref{anapole_moment2}), the dimensionless parameter $\kappa_a$ in Eq.~(\ref{anapole_EM_Hamiltonian}) is given by: 
\begin{equation}
\label{kappa_parameter1}
\kappa_a \approx - \frac{g^A_{N'} g^V_N}{G_F m_{Z'}^2} \frac{9 \sqrt{2} \alpha \mu }{10 m r_0 } A^{2/3} ~~\textrm{for $m_{Z'} \gg 1/r_0$} \, ,
\end{equation}
\begin{equation}
\label{kappa_parameter2}
\kappa_a \approx - \frac{g^A_{N'} g^V_N}{G_F} \frac{9 \sqrt{2} \alpha \mu r_0}{25 m } A^{4/3} ~~\textrm{for $m_{Z'} \ll 1/R$} \, ,
\end{equation}
where $\alpha = e^2 \approx 1/137$ is the electromagnetic fine-structure constant. 

The interaction (\ref{anapole_EM_Hamiltonian}) induces nuclear-spin-dependent PNC effects in atoms and molecules, allowing the determination of the parameter $\kappa_a$. 
The only successful measurement of a nuclear anapole moment to date was performed in Ref.~\cite{Wood1997}. 
The experimentally measured value of $\kappa_a$ for the $^{133}$Cs nucleus is \cite{Wood1997,Murray1997}:
\begin{equation}
\label{kappa_exp}
\kappa_a = 0.364(62) \, . 
\end{equation}
Single-particle nuclear shell-model calculations of $\kappa_a$ for the $^{133}$Cs nucleus have been performed in Refs.~\cite{Flambaum1984,Murray1997} using nucleon interaction constants from Ref.~\cite{Holstein1980}, while many-body corrections have been considered in Refs.~\cite{Korov1994,Dmitriev1997,Dmitriev2000,Haxton2001,Haxton2002}. 
For consistency with the single-particle approach adopted in the present work, we likewise use the results of single-particle calculations \cite{Murray1997}: 
\begin{equation}
\label{kappa_theor}
\kappa_a = 0.27(8) \, . 
\end{equation}

Comparing the measured and predicted values of $\kappa_a$ in Eqs.~(\ref{kappa_exp}) and (\ref{kappa_theor}), and using expressions (\ref{kappa_parameter1}) and (\ref{kappa_parameter2}), we place the following constraints on the interaction parameters in Eq.~(\ref{general_formula_SP}): 
\begin{equation}
\label{proton-nucleon_limit1}
\frac{\left| g^A_{p} g^V_N \right|}{m_{Z'}^2} < 2.3 \times 10^{-5}~\textrm{GeV}^{-2} ~~\textrm{for $m_{Z'} \gg 1/r_0$} \, , 
\end{equation}
\begin{equation}
\label{proton-nucleon_limit2}
\left| g^A_{p} g^V_N \right| < 6.0 \times 10^{-8} ~~\textrm{for $m_{Z'} \ll 1/R$} \, . 
\end{equation}

\begin{figure}[h!]
\begin{center}
\includegraphics[width=8.5cm]{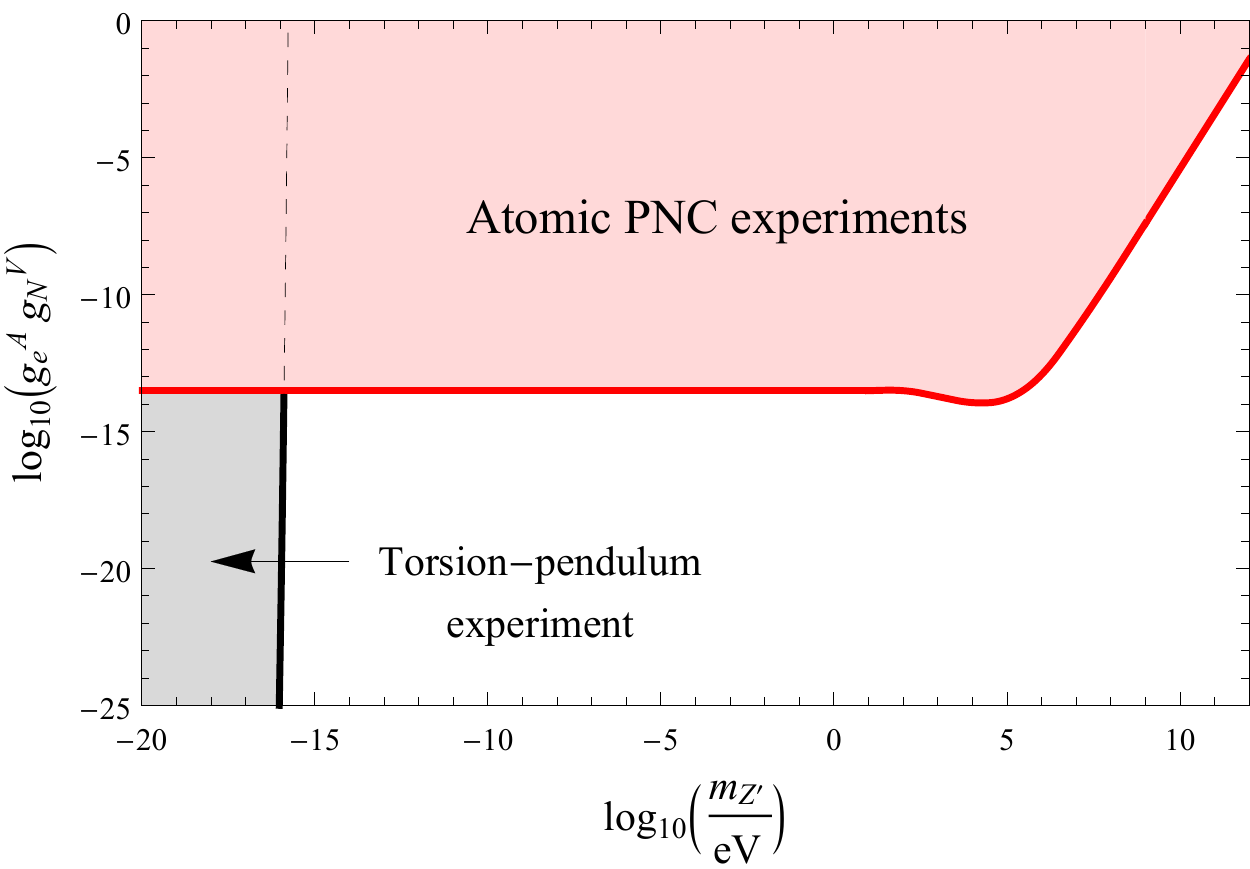} 
\includegraphics[width=8.5cm]{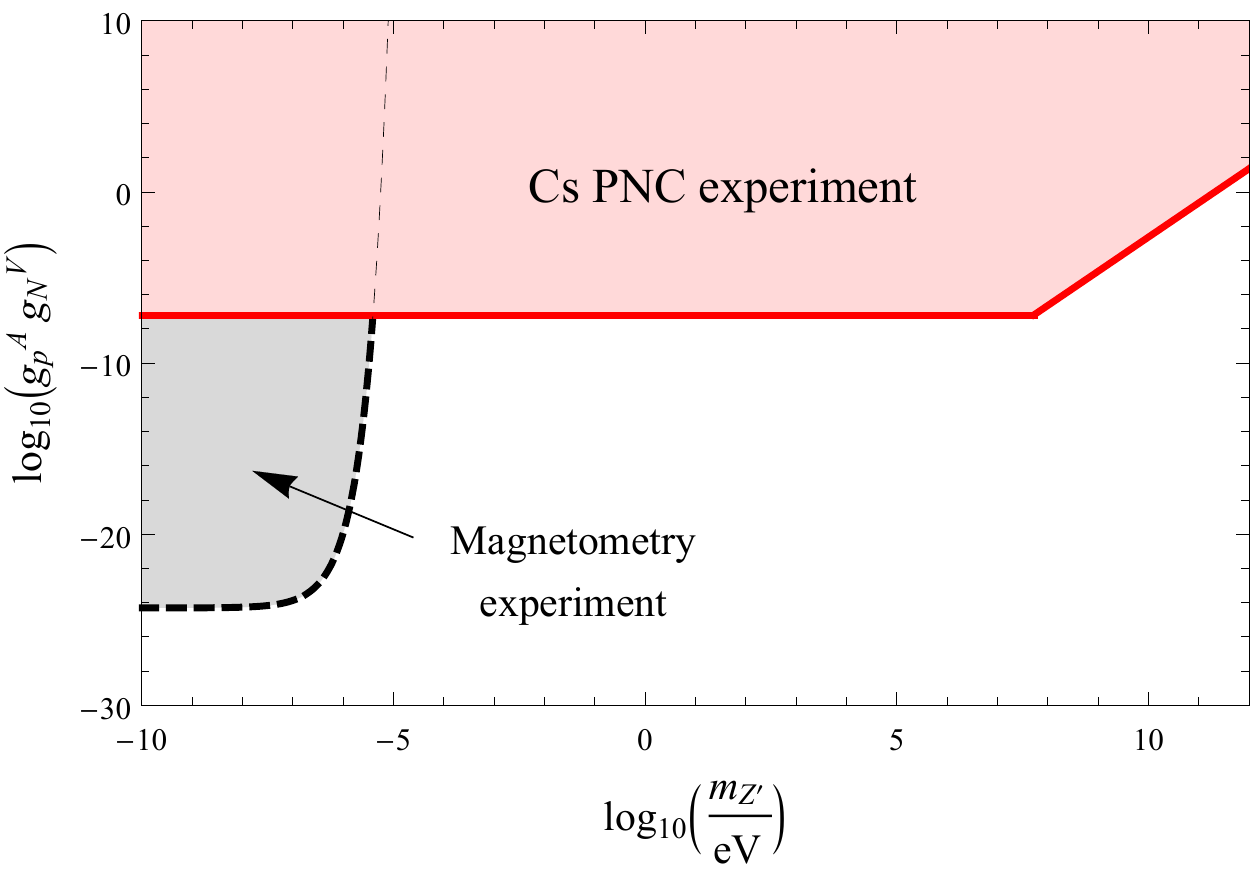} 
\caption{(Color online) 
Limits on the \emph{P}-violating vector-pseudovector nucleon-electron (top figure) and nucleon-proton (bottom figure) interactions mediated by a generic vector boson of mass $m_{Z'}$, as defined in Eq.~(\ref{general_formula_SP}). 
The regions in red correspond to regions of parameters excluded by the present work from consideration of atomic parity nonconservation experiments. 
The regions in grey correspond to existing constraints from torsion-pendulum and magnetometry experiments \cite{Adelberger2006,Adelberger2008,Romalis2009}. 
} 
\label{fig:Results_summary}
\end{center}
\end{figure}

\textbf{Conclusions.} --- 
We have derived limits on the \emph{P}-violating vector-pseudovector nucleon-electron and nucleon-proton interactions mediated by a generic vector boson of arbitrary mass from atomic PNC experiments (see Table~\ref{tab:table2} for a summary of limits). 
Our derived limits on the electron-nucleon interaction improve on existing bounds from torsion-pendulum experiments \cite{Adelberger2006,Adelberger2008} by many orders of magnitude for $m_{Z'} \gtrsim 10^{-16}~\textrm{eV}$ (see Fig.~\ref{fig:Results_summary}). 
For non-isotopically-invariant interactions of a vector boson with nucleons (i.e., the vector boson couples with different strengths to the proton and neutron), our constraints on the proton-nucleon interaction are complementary to existing bounds from magnetometry experiments on the neutron-nucleon interaction \cite{Romalis2009}, while for isotopically-invariant interactions of a vector boson with nucleons, our derived limits on the nucleon-nucleon interaction improve on existing bounds from magnetometry experiments \cite{Romalis2009} by many orders of magnitude for $m_{Z'} \gtrsim 10^{-5}~\textrm{eV}$ (see Fig.~\ref{fig:Results_summary}). 

Ongoing and future PNC experiments with atoms \cite{Williams2013Ba+PNC,Leefer2014DyPNC,Antypas2017YbPNC,Tandecki2014FrPNC,Portela2013Ra+PNC} and molecules \cite{Cahn2014BaFpnc,Isaev2010RaFpnc} may improve on the level of sensitivity demonstrated in the present work. 
In particular, atomic PNC experiments that involve the mixing of atomic states of high angular momentum (e.g., in Dy \cite{Leefer2014DyPNC}) may be particularly sensitive to the electron-nucleon interaction mediated by a low-mass vector boson, since a large centrifugal barrier does not necessarily suppress PNC effects in this case (in contrast to the case of a high-mass vector boson, where the effects arise mainly in the vicinity of the atomic nucleus). 
Molecular PNC experiments are primarily sensitive to nuclear-spin-dependent PNC effects \cite{Flambaum1978,Kozlov1995}, and thus may provide improved sensitivity to nucleon-nucleon interactions.

\textbf{Acknowledgements.} --- 
This work was supported in part by the Australian Research Council. 
V.~V.~F.~was supported by the Gutenberg Research College Fellowship. 
Y.~V.~S.~was supported by the Humboldt Research Fellowship.




\begin{thebibliography}{99} 

\bibitem{Yang1956} T.~D.~Lee and C.~N.~Yang, Phys.~Rev.~\textbf{104}, 254 (1956); Phys.~Rev.~\textbf{106}, 1371(E) (1957). 

\bibitem{Wu1957} C.~S.~Wu, E.~Ambler, R.~W.~Hayward, D.~D.~Hoppes, R.~P.~Hudson, Phys.~Rev.~\textbf{105}, 1413 (1957). 

\bibitem{Khriplovich1991Book} I.~B.~Khriplovich, \emph{Parity Nonconservation in Atomic Phenomena}, Gordon and Breach, Philadelphia, 1991. 

\bibitem{Ginges2004Review} J.~S.~M.~Ginges and V.~V.~Flambaum, Phys.~Rept.~\textbf{397}, 63 (2004). 

\bibitem{Roberts2015ReviewPNC} B.~M.~Roberts, V.~A.~Dzuba, V.~V.~Flambaum, Annu.~Rev.~Nucl.~Part.~Sci.~\textbf{65}, 63 (2015). 

\bibitem{Bouchiat1974} M.-A.~Bouchiat and C.~Bouchiat, Le Journal de Physique~\textbf{35}, 899 (1974). 

\bibitem{Bouchiat1982} M.-A.~Bouchiat, J.~Guena, L.~Hunter, L.~Pottier, Phys.~Lett.~B~\textbf{117}, 358 (1982). 

\bibitem{Sushkov1989} V.~A.~Dzuba, V.~V.~Flambaum, O.~P.~Sushkov, Phys.~Lett.~A~\textbf{141}, 147 (1989). 

\bibitem{Johnson1992} S.~A.~Blundell, J.~Sapirstein, W.~R.~Johnson, Phys.~Rev.~D~\textbf{45}, 1602 (1992). 

\bibitem{Wood1997} C.~S.~Wood \emph{et al}., Science~\textbf{275}, 1759 (1997). 

\bibitem{Kozlov2001} M.~G.~Kozlov, S.~G.~Porsev, I.~I.~Tupitsyn, Phys.~Rev.~Lett.~\textbf{86}, 3260 (2001). 

\bibitem{Safronova2002} A.~A.~Vasilyev, I.~M.~Savukov, M.~S.~Safronova, H.~G.~Berry, Phys.~Rev.~A~\textbf{66}, 020101 (2002). 

\bibitem{Ginges2002} V.~A.~Dzuba, V.~V.~Flambaum, J.~S.~M.~Ginges, Phys.~Rev.~D~\textbf{66}, 076013 (2002). 

\bibitem{Shabaev2005} V.~M.~Shabaev, I.~I.~Tupitsyn, K.~Pachucki, G.~Plunien, V.~A.~Yerokhin, Phys.~Rev.~A~\textbf{72}, 062105 (2005). 

\bibitem{Derevianko2009} S.~G.~Porsev, K.~Beloy, A.~Derevianko, Phys.~Rev.~Lett.~\textbf{102}, 181601 (2009). 

\bibitem{Roberts2012} V.~A.~Dzuba, J.~C.~Berengut, V.~V.~Flambaum, B.~Roberts, Phys.~Rev.~Lett.~\textbf{109}, 203003 (2012). 

\bibitem{Zeldovich1957} Ya.~B.~Zel'dovich, Zh.~Eksp.~Teor.~Fiz.~\textbf{33}, 1531 (1957); [Sov.~Phys.~JETP~\textbf{6}, 1184 (1958)]. 

\bibitem{Flambaum1980} V.~V.~Flambaum and I.~B.~Khriplovich, Zh.~Eksp.~Teor.~Fiz.~\textbf{79}, 1656 (1980); [Sov.~Phys.~JETP~\textbf{52}, 835 (1980)]. 

\bibitem{Flambaum1984} V.~V.~Flambaum, I.~B.~Khriplovich, O.~P.~Sushkov, Phys.~Lett.~B~\textbf{146}, 367 (1984). 

\bibitem{Murray1997} V.~V.~Flambaum, D.~W.~Murray, Phys.~Rev.~C~\textbf{56}, 1641 (1997). 

\bibitem{Bouchiat1983} C.~Bouchiat and C.~A.~Piketty, Phys.~Lett.~B~\textbf{128}, 73 (1983). 

\bibitem{Marciano1990} W.~J.~Marciano and J.~L.~Rosner, Phys.~Rev.~Lett.~\textbf{65}, 2963 (1990); Phys.~Rev.~Lett.~\textbf{68}, 898 (1992). 

\bibitem{Bouchiat2004} C.~Bouchiat and P.~Fayet, Phys.~Lett.~B~\textbf{608}, 87 (2005). 

\bibitem{Marciano2012a} H.~Davoudiasl, H.-S.~Lee, W.~J.~Marciano, Phys.~Rev.~D~\textbf{85}, 115019 (2012). 

\bibitem{Marciano2012b} H.~Davoudiasl, H.-S.~Lee, W.~J.~Marciano, Phys.~Rev.~Lett.~\textbf{109}, 031802 (2012). 

\bibitem{Stadnik2014a} Y.~V.~Stadnik and V.~V.~Flambaum, Phys.~Rev.~D~\textbf{89}, 043522 (2014). 

\bibitem{Stadnik2014b} B.~M.~Roberts, Y.~V.~Stadnik, V.~A.~Dzuba, V.~V.~Flambaum, N.~Leefer, D.~Budker, Phys.~Rev.~Lett.~\textbf{113}, 081601 (2014). 

\bibitem{Stadnik2014c} B.~M.~Roberts, Y.~V.~Stadnik, V.~A.~Dzuba, V.~V.~Flambaum, N.~Leefer, D.~Budker, Phys.~Rev.~D~\textbf{90}, 096005 (2014). 

\bibitem{Footnote1} We consider a $Z'$-boson that does not kinetically mix with the $Z$-boson of the SM. 


\bibitem{s-d_PNC2001} V.~A.~Dzuba, V.~V.~Flambaum, J.~S.~M.~Ginges, Phys.~Rev.~A~\textbf{63}, 062101 (2001). 

\bibitem{s-d_PNC2011} V.~A.~Dzuba and V.~V.~Flambaum, Phys.~Rev.~A~\textbf{83}, 052513 (2011). 

\bibitem{Dzuba2011YbPNC} V.~A.~Dzuba and V.~V.~Flambaum, Phys.~Rev.~A~\textbf{83}, 042514 (2011). 


\bibitem{Dzuba1987corr} V.~A.~Dzuba, V.~V.~Flambaum, P.~G.~Silvestrov, O.~P.~Sushkov, J.~Phys.~B~\textbf{20}, 1399 (1987). 

\bibitem{Dzuba1989corr} V.~A.~Dzuba, V.~V.~Flambaum, O.~P.~Sushkov, Phys.~Lett.~A~\textbf{140}, 493 (1989). 


\bibitem{Tsigutkin2009YbPNC} K.~Tsigutkin, D.~R.~Dounas-Frazer, A.~Family, J.~E.~Stalnaker, V.~V.~Yashchuk, D.~Budker, Phys.~Rev.~Lett.~\textbf{103}, 071601 (2009). 

\bibitem{Fortson1995TlPNC} P.~A.~Vetter, D.~M.~Meekhof, P.~K.~Majumder, S.~K.~Lamoreaux, E.~N.~Fortson, Phys.~Rev.~Lett.~\textbf{74}, 2658 (1995). 

\bibitem{Kozlov2001TlPNC} M.~G.~Kozlov, S.~G.~Porsev, W.~R.~Johnson, Phys.~Rev.~A~\textbf{64}, 052107 (2001). 


\bibitem{Holstein1980} B.~Desplanques, J.~F.~Donoghue, B.~R.~Holstein, Annals of Physics~\textbf{124}, 449 (1980). 

\bibitem{Korov1994} V.~V.~Flambaum and O.~K.~Vorov, Phys.~Rev.~C~\textbf{49}, 1827 (1994). 

\bibitem{Dmitriev1997} V.~F.~Dmitriev and V.~B.~Telitsin, Nucl.~Phys.~A~\textbf{613}, 237 (1997). 

\bibitem{Dmitriev2000} V.~F.~Dmitriev and V.~B.~Telitsin, Nucl.~Phys.~A~\textbf{674}, 168 (2000). 

\bibitem{Haxton2001} W.~C.~Haxton, C.-P.~Liu, M.~J.~Ramsey-Musolf, Phys.~Rev.~Lett.~\textbf{86}, 5247 (2001). 

\bibitem{Haxton2002} W.~C.~Haxton, C.-P.~Liu, M.~J.~Ramsey-Musolf, Phys.~Rev.~C~\textbf{65}, 045502 (2002). 


\bibitem{Adelberger2006} B.~R.~Heckel, C.~E.~Cramer, T.~S.~Cook, E.~G.~Adelberger, S.~Schlamminger, U.~Schmidt, Phys.~Rev.~Lett.~\textbf{97}, 021603 (2006). 

\bibitem{Adelberger2008} B.~R.~Heckel, E.~G.~Adelberger, C.~E.~Cramer, T.~S.~Cook, S.~Schlamminger, U.~Schmidt, Phys.~Rev.~D~\textbf{78}, 092006 (2008). 

\bibitem{Romalis2009} G.~Vasilakis, J.~M.~Brown, T.~W.~Kornack, M.~V.~Romalis, Phys.~Rev.~Lett.~\textbf{103}, 261801 (2009). 


\bibitem{Williams2013Ba+PNC} S.~R.~Williams \emph{et al}., Phys.~Rev.~A~\textbf{88}, 012515 (2013). 

\bibitem{Leefer2014DyPNC} N.~Leefer, L.~Bougas, D.~Antypas, D.~Budker, arXiv:1412.1245. 

\bibitem{Antypas2017YbPNC} D.~Antypas, A.~Fabricant, L.~Bougas, K.~Tsigutkin, D.~Budker, Hyperfine Interact.~\textbf{238}, 21 (2017). 

\bibitem{Tandecki2014FrPNC} M.~Tandecki \emph{et al}., J.~Instrum.~\textbf{9}, P10013 (2014). 

\bibitem{Portela2013Ra+PNC} M.~Nunez Portela \emph{et al}., Appl.~Phys.~B~\textbf{114}, 173 (2013). 

\bibitem{Cahn2014BaFpnc} S.~B.~Cahn \emph{et al}., Phys.~Rev.~Lett.~\textbf{112}, 163002 (2014). 

\bibitem{Isaev2010RaFpnc} T.~A.~Isaev, S.~Hoekstra, R.~Berger, Phys.~Rev.~A~\textbf{82}, 052521 (2010). 


\bibitem{Flambaum1978} O.~P.~Sushkov and V.~V.~Flambaum, Zh.~Eksp.~Teor.~Fiz.~\textbf{75}, 1208 (1978); [Sov.~Phys.~JETP~\textbf{48}, 608 (1978)]. 

\bibitem{Kozlov1995} M.~G.~Kozlov and L.~N.~Labzowsky, J.~Phys.~B~\textbf{28}, 1933 (1995). 




\end{thebibliography}
\end{document}